\documentclass{emulateapj}
\lefthead{Adelberger et al.}

\righthead{Clustering of UV-selected Galaxies at $z\sim 2$}

\submitted{Received 2004 September 30; Accepted 2005 January 7}

\begin{document}
\newcommand{\lya}{Lyman~$\alpha$}
\newcommand{\lyb}{Lyman~$\beta$}
\newcommand{\degpoint}{\mbox{$^\circ\mskip-7.0mu.\,$}}
\newcommand{\minpoint}{\mbox{$'\mskip-4.7mu.\mskip0.8mu$}}
\newcommand{\secpoint}{\mbox{$''\mskip-7.6mu.\,$}}
\newcommand{\sqdeg}{\mbox{${\rm deg}^2$}}
\newcommand{\squig}{\sim\!\!}
\newcommand{\subsun}{\mbox{$_{\twelvesy\odot}$}}
\newcommand{\et}{{\it et al.}~}
\newcommand{\Rs}{{\cal R}}

\def\ltsima{$\; \buildrel < \over \sim \;$}
\def\simlt{\lower.5ex\hbox{\ltsima}}
\def\gtsima{$\; \buildrel > \over \sim \;$}
\def\simgt{\lower.5ex\hbox{\gtsima}}
\def\propsima{$\; \buildrel \propto \over \sim \;$}
\def\simprop{\lower.5ex\hbox{\propsima}}
\def\arcs{$''~$}
\def\arcm{$'~$}

\title{STRONG SPATIAL CLUSTERING OF UV-SELECTED GALAXIES WITH MAGNITUDE $K_s<20.5$ AND REDSHIFT $Z\sim 2$\altaffilmark{1}}

\author{\sc Kurt L. Adelberger\altaffilmark{2}}
\affil{Carnegie Observatories, 813 Santa Barbara St., Pasadena, CA, 91101}

\author{\sc Dawn K. Erb, Charles C. Steidel, and Naveen A. Reddy}
\affil{Palomar Observatory, Caltech 105--24, Pasadena, CA 91125}

\author{\sc Max Pettini}
\affil{Institute of Astronomy, Madingley Road, Cambridge CB3 0HA, UK}
                                                                                
\author{\sc Alice E. Shapley\altaffilmark{3}}
\affil{University of California, Department of Astronomy, 601 Campbell Hall, Berkeley, CA 94720}

\altaffiltext{1}{Based, in part, on data obtained at the W.M. Keck
Observatory, which is operated as a scientific partnership between
the California Institute of Technology, the University of California,
and NASA, and was made possible by the generous financial support
of the W.M. Keck Foundation.}
\altaffiltext{2}{Carnegie Fellow}
\altaffiltext{3}{Miller Fellow}

\begin{abstract}
We obtained deep $8.5'\times 8.5'$ near-infrared images within
four high-redshift survey fields, measured the $K_s$ magnitudes
of 300 optically selected galaxies with spectroscopic redshift
$1.8\simlt z\simlt 2.6$,
and compared the spatial clustering strength of 
galaxies with $K_s<20.5$ and $K_s>20.5$.
We found at
greater than $95$\% confidence that the brighter galaxies cluster more strongly.
The best-fit correlation lengths for the
bright and faint samples are $10\pm 3$ and $4\pm 0.8 h^{-1}$
comoving Mpc, respectively ($1\sigma$),
although the unusual density of bright QSOs in one of our survey fields
may imply that the result is not representative of the universe as a whole.
Neglecting this possibility,
the correlation length for the optically selected sample with $K_s<20.5$
agrees well with that reported for comparably bright
near-IR-selected samples.   The differences in correlation length
between optically selected and near-IR-selected samples
have been presented as evidence that the two techniques find
orthogonal populations of high-redshift galaxies.
Our results favor a more nuanced view.
\end{abstract}

\section{INTRODUCTION}
\label{sec:intro}
Near-infrared (IR) surveys of the high-redshift universe 
(Cimatti et al. 2002, Franx et al. 2003, Abraham et al. 2004)
were designed to find passively evolving
galaxies that had been missed by optical surveys,
but they also uncovered an intriguing population of 
galaxies
whose star-formation rates $\dot M_\ast\simgt 200M_\odot$ yr$^{-1}$ and
stellar masses $M_\ast\simgt 2\times 10^{11}M_\odot$
dwarfed those of optically-selected galaxies.
Since these star-forming
galaxies have the properties 
expected for the progenitors of giant elliptical galaxies,
many researchers have begun to assume that near-IR surveys find the
progenitors of ellipticals, that optical surveys find the progenitors
of less massive local galaxies, and (by extension) that the high-redshift
universe is divided into two distinct star-forming populations with divergent
evolutionary paths.

Although some of the near-IR-selected galaxies
clearly do not satisfy common optical selection criteria
(see, e.g., Daddi et al. 2004; Forster-Schreiber et al. 2004;
van Dokkum et al. 2004),
the evidence for two distinct star-forming populations
still leaves room for doubt.  First, the differences
in mean stellar masses and star-formation rates
might simply reflect the fact that near-IR surveys
tend to be shallower than optical surveys, detecting
only the brightest tip of the luminosity distribution.
Since the
near-IR-selected star-forming galaxies are less numerous than optically selected galaxies,
a significant fraction of them 
could be present in optical surveys 
without
changing the mean properties of optical sources
in an appreciable way.
The unusual measured properties of bright near-IR galaxies do not
imply that they are absent from optical surveys any more than
the high mean income of Luxembourgers
implies that they are a previously undiscovered population in the E.~U.
Second, sub-millimeter surveys
show that the majority
of the most rapidly star-forming galaxies at high redshift
are bright enough and blue enough to be included in existing optical surveys
(e.g., Chapman et al. 2005; see their Figure~6).
Finally, there are physical reasons to believe that at least some
of the elliptical galaxies in the local universe descended from
the objects detected in optical surveys.  Ellipticals are the only
local galaxies that have the spatial distribution expected
for the descendants of optically selected star-forming galaxies
at $z>1.5$, for example
(see Adelberger et al. 2005).  

Partly in order to help 
clarify the connection between optical and near-IR
populations,
our group has obtained $K_s$ images throughout the
course of its optical surveys at $1.4<z<3.5$. 
Recent analyses
of the $K_s$ images (Erb et al. 2005, in preparation) reveal that many
optically-selected high-redshift galaxies 
have the magnitudes $K_s<20$ characteristic of the supposedly distinct
bright near-IR-selected populations.
Shapley et al. (2004) showed 
that these optically-selected galaxies with $K_s<20$
have star-formation rates, stellar masses, and metallicities
similar to those of star-forming galaxies in near-IR-selected surveys
to $K_s=20$.  This suggested that there might be substantial overlap between
near-IR and optical populations at high redshift,
a conclusion
reinforced by their estimate
that the number density of high-redshift galaxies with $K_s<20$
in optically-selected surveys is at least half as large as
the number density reported by near-IR surveys.
Daddi et al. (2004, 2005) reached a similar conclusion
from a different starting point.

The present paper extends the work of Shapley et al. (2004) in a small
way by showing that optically-selected galaxies at $z\sim 2$
with bright $K_s$ magnitudes cluster as strongly as their
near-IR-selected counterparts.  This suggests that the $K_s$-bright
star-forming galaxies detected in optical surveys may not be
enormously different from the $K_s$-bright star-forming galaxies
that are missed.
We will discuss the implications further
in~\S~\ref{sec:summary}, after first describing our
data in~\S~\ref{sec:data}, our method for measuring $r_0$
in~\S~\ref{sec:methods}, and our raw results
in~\S~\ref{sec:results}.

\section{DATA}
\label{sec:data}
\begin{figure}
\plotone{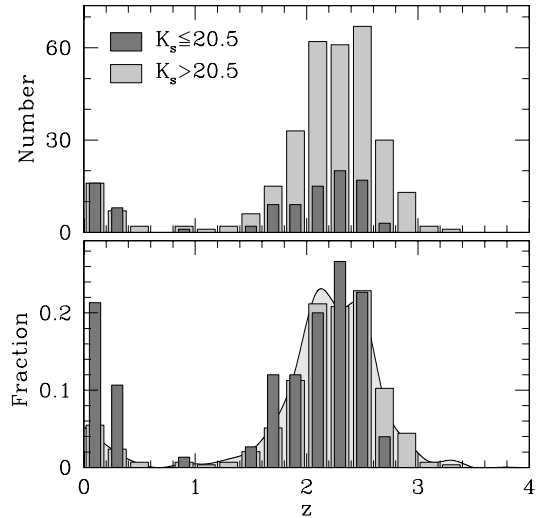}
\caption{
Redshift distributions for the galaxies in our samples.  The top
panel shows the raw number of galaxies with $K_s\leq 20.5$ and
$K_s>20.5$ (Vega system).  The bottom panel shows the same data normalized to
constant area for $z>1$.  The overall redshift distribution
for the ``BX'' survey of Steidel et al. (2004) is shown as a smooth
curve in the background.  The low redshift tail with $z<1$ is
excluded from the clustering analysis of this paper.
\label{fig:nz}
}
\end{figure}

The galaxies we studied lie within the
fields GOODS-N, Q1623, Q1700, and Q2343 of the spectroscopic survey of
Steidel et al. (2004).  Steidel et al. (2004) obtained
$U_nG{\cal R}$ images of these fields, constructed a catalog
of objects with AB-magnitude
${\cal R}\leq 25.5$
whose colors satisfied
the $z\sim 2$ ``BX'' selection criteria of Adelberger et al. (2004),
measured spectroscopic redshifts for several hundred of the
objects in this catalog, and, after excluding
objects with $z<1$,  found 
that the mean redshift and r.m.s. dispersion
$\bar z\pm \sigma_z = 2.21\pm 0.34$
agreed well with the range targeted by the photometric selection criteria.
To measure the near-IR magnitudes of some of the galaxies in the catalog,
we (Erb et al. 2005, in preparation)
subsequently used the Wide-Field Infrared Camera (WIRC; Wilson et al. 2003)
on the Hale 5m telescope to obtain a deep $K_s$ image
($\sim 11$hr integration
time, $5\sigma$ point-source detection threshold of 
$K_s\simeq 22.3$ in the Vega system)
of a roughly $8.5'\times 8.5'$ region within the larger
optically-observed area in each of the four fields.
A total of 1598 objects in the photometric BX catalogs
of Steidel et al. (2004) fell within the areas observed with
WIRC.  We restrict our analysis to the 368 of these sources that have
measured redshifts.
Most (300) of the sources with redshifts were detected in $K_s$ at $3\sigma$
significance or better, and 75 were brighter than the $K_s=20.5$
threshold that we use below to divide our sample into bright
and faint subsamples.
See table~\ref{tab:fields}; all near-IR magnitudes here and
elsewhere are in the Vega system, and all optical magnitudes
are AB.
$K_s=20.5$ was chosen because it was the brightest threshold that
left us with enough objects in the bright subsample; only 20
objects satisfy the threshold $K_s=20$.
The redshift distributions for the bright and faint subsamples
are shown in figure~\ref{fig:nz}.  

\begin{deluxetable}{lccccc
}\tablewidth{0pc}
\scriptsize
\tablecaption{Observed fields}
\tablehead{
        \colhead{Name} &
        \colhead{$\Delta\Omega_{\rm opt}$\tablenotemark{a}} &
        \colhead{$\Delta\Omega_{\rm IR}$\tablenotemark{b}} &
        \colhead{$N_{\rm img}^{\rm IR}$\tablenotemark{c}} &
        \colhead{$N_{\rm det}^{\rm IR}$\tablenotemark{d}} &
        \colhead{$N_{<20.5}^{\rm IR}$\tablenotemark{e}} 
}
\startdata
GOODS-N & $14.0\times 13.4$ & $8.7\times 8.0$  &  84  &  60  & 10 \\
Q1623   & $16.1\times 11.6$ & $8.7\times 8.8$  & 121  &  98  & 30 \\
Q1700   & $11.5\times 11.0$ & $8.5\times 7.8$  &  62  &  57  & 13 \\
Q2343   & $22.5\times 8.5$  & $8.7\times 5.4$  & 101  &  85  & 22 \\
total   & $695$             & $259$            & 368  & 300  & 75 \\
\enddata
\tablenotetext{a}{Area imaged in $U_nG{\cal R}$ and observed spectroscopically, square arcmin}
\tablenotetext{b}{Area imaged in $K_s$ and observed spectroscopically, square arcmin}
\tablenotetext{c}{Number of sources in the BX catalog that fall within the $K_s$ image and have spectroscopic redshifts $z>1$.}
\tablenotetext{d}{Number of sources in the BX catalog that are detected in the $K_s$ image at $3\sigma$ confidence and have spectroscopic redshifts $z>1$.}
\tablenotetext{e}{Number of sources in the BX catalog with $K_s\leq 20.5$ (Vega system) and spectroscopic redshifts $z>1$.}
\label{tab:fields}
\end{deluxetable}

The total number of galaxies in our subsamples would be significantly
higher if our spectroscopy were 100\% complete.  Since the 
fraction of spectroscopically observed BX candidates that have $1.8<z<2.8$ is
roughly $0.8$, $0.65$, $0.5$ for $K_s>20.5$, $K_s\leq 20.5$, $K_s\leq 20$,
we estimate from the apparent magnitude distribution of all BX candidates
that the total number of BX candidates in our fields with $1.8<z<2.8$
is 1000, 200, 80 for the same three magnitude ranges.
Although our spectroscopy is far from complete,
our spectroscopic catalog of galaxies with $K_s<20.5$ is already
about an order of magnitude larger than any near-IR selected catalog
in the same redshift range.

\section{METHODS}
\label{sec:methods}
We estimated the correlation lengths of the samples with two
approaches.  Both were based on counting the observed number of galaxy pairs
with comoving radial separation $|\Delta Z|<\ell$ and comparing
to the number expected for a reduced correlation function
of the form $\xi(r) = (r/r_0)^{-\gamma}$.  Letting $Z_i$ denote the comoving
distance to redshift $z_i$ and adopting the shorthand $Z_{ij}\equiv Z_i-Z_j$,
we counted
the observed number of pairs in the each field with radial separation $0\leq|\Delta Z|<\ell$
and summed over fields to get
$N_{\rm obs}(0,\ell)$, the total number of pairs with $0\leq|\Delta Z|<\ell$.

If the selection function $P(z)$ is accurately known, 
the expected number of pairs with $0\leq |\Delta Z|<\ell$
can be estimated for any assumed value of $r_0$.
Suppose we know that the $j$th galaxy in a field has redshift $z_j$
and that the $i$th galaxy lies at an angular separation
of $\theta_{ij}$.  Then the probability that 
the unknown redshift $z_i$ will satisfy $0\leq |Z_{ij}|<\ell$ is
(Adelberger 2005)
\begin{eqnarray}
P(0\leq |Z_{ij}|<\ell\,|\,z_j\theta_{ij}) = \frac{P(z_j) [2\ell g^{-1}(z_j)+a{\cal I}]}{1+aP(z_j)}
\label{eq:kpair1}
\end{eqnarray}
where
$a\equiv r_0^\gamma[f(z_j)\theta_{ij}]^{1-\gamma}g^{-1}(z_j)\beta(\gamma),$
$g(z)\equiv c/H(z)$ is the change in comoving distance with redshift,
$f(z)\equiv (1+z)D_A(z)$ is the change in comoving distance with angle,
$D_A(z)$ is the angular diameter distance,
$\beta(\gamma)\equiv B[1/2,(\gamma-1)/2]$,
$B$ is the beta function in the convention of Press et al. (1992),
and
${\cal I}$ is related to the incomplete beta function $I_x$
of Press et al. (1992) through
${\cal I} \equiv I_x[1/2,(\gamma-1)/2]$
with
$x\equiv \ell^2 / \{\ell^2+[f(z_j)\theta_{ij}]^2\}$.
The expected number of pairs with
$0\leq |\Delta Z|<\ell$ is equal to the sum of
the probabilities $P(0\leq |Z_{ij}|<\ell\,|\,z_j\theta_j)$ for
each unique pair,
\begin{equation}
N_{\rm exp} = \frac{1}{2}\sum_{i\neq j}^{\rm pairs} P(0\leq |Z_{ij}|<\ell\, |\, z_j\theta_{ij}),
\label{eq:kpair}
\end{equation}
where the sum includes only pairs in which both galaxies lie in the same field.

Our first approach towards estimating $r_0$ is to find the value
that makes the expected number of pairs
$N_{\rm exp}(0,\ell)$ equal the observed number $N_{\rm obs}(0,\ell)$. 
By summing over the pair counts in all fields when calculating
$N_{\rm exp}$ and $N_{\rm obs}$, we are implicitly assuming that
the selection function $P(z)$ does not vary appreciably from one field
to the next.  The estimated values of $r_0$ will be incorrect
if this is not the case, but the analysis is otherwise insensitive
to variations in data quality from field to field.

Even if the selection function is reasonably constant among the fields,
as it probably is, the inferred value of $r_0$ 
will still depend sensitively on its
assumed shape (Adelberger 2005).  
This is the major weakness of equation~\ref{eq:kpair}.
An alternative is to remove the sensitivity to the selection function by
dividing the number of pairs with $0\leq \Delta Z < \ell$
by the number with $0\leq \Delta Z < 2\ell$.  As shown by
Adelberger et al. (2005), the expectation value of this quotient
is 
\begin{equation}
\Biggl\langle\frac{N_{\rm obs}(0,\ell)}{N_{\rm obs}(0,2\ell)}\Biggr\rangle \simeq \frac{\sum_{i\neq j}\int_0^\ell dZ [1+\xi(R_{ij},Z)]}{\sum_{i\neq j}\int_0^{2\ell} dZ [1+\xi(R_{ij},Z)]},
\label{eq:kdaddi}
\end{equation}
where $\xi(R,Z)\equiv \xi[(R^2+Z^2)^{1/2}]$
and $R_{ij}\equiv (1+z_i)D_A(z_i) \theta_{ij}$,
regardless of the selection function, as long as
the angular diameter distance changes slowly with redshift,
$\ell$ is significantly larger than the comoving uncertainty in
a galaxy's radial position, 
$2\ell$ is significantly smaller than the selection function's
width, and $N_{\rm obs}(0,2\ell)$ is large enough that
$\langle 1/N_{\rm obs}(0,2\ell)\rangle \simeq 1 / \langle N_{\rm obs}(0,2\ell) \rangle$.   
Our second approach to estimating $r_0$ is to find
the value that makes the right-hand side of
equation~\ref{eq:kdaddi} equal the observed quantity
$N_{\rm obs}(0,\ell)/N_{\rm obs}(0,2\ell)$.  The estimate
of $r_0$ from this approach is barely affected by even
major changes in data quality from one field to the next.

Although this insensitivity is a significant
advantage, the disadvantage of the second approach is
that it can be noisy.
We adopt both approaches because one 
is less susceptible to systematic errors
and the other is less susceptible to random.
See Adelberger (2005) for further discussion.

In both approaches we adopt $\ell=20h^{-1}$ Mpc, considerably
larger than the biggest plausible peculiar velocity or redshift measurement
error, and assume that the correlation function has
the slope of $\gamma=1.6$ measured from a larger sample of similar
galaxies by Adelberger et al. (2005).  Adopting $\gamma=1.8$ instead
would lower our derived correlation lengths by less than 10\%.
Eliminating pairs with $\theta_{ij}>300''$ (e.g., Adelberger 2005)
does not significantly alter the results.

\section{RESULTS}
\label{sec:results}
Equations~\ref{eq:kpair} and~\ref{eq:kdaddi} lead
to the estimates $r_0 = 3.8$, $4.4h^{-1}$ comoving Mpc
for the subsample with $K_s>20.5$ and
and $r_0 = 9.9$, $10.8h^{-1}$ comoving Mpc
for the subsample with $K_s\leq 20.5$.
To calculate the significance of the observed difference between
the two samples' correlation lengths, we split our full 368-object sample
into two parts at random, rather than according to $K_s$ magnitude,
measured the difference in $r_0$ between the two parts, then repeated
the process over and over, with a different random splitting
each time, to estimate the distribution of
$\Delta r_0$ under the null hypothesis that $r_0$ is unrelated
to $K_s$ magnitude.  Each random splitting placed 75 of our
368 objects in one subsample and 293 in another, mimicking the
75/293 division of the actual splitting at $K_s=20.5$.
We observed a difference in $r_0$ as large as the observed
difference in fewer than $5$\% of the randomized catalogs,
implying that the observed difference is significant
at better than 95\% confidence.

We used two approaches to estimate
the size of the uncertainties in the
correlation lengths for the two samples.
First, we measured the dispersion in the
the best-fit value of $r_0$ from each individual field.
Although in principle this empirical approach should work well,
in practice $r_0$ is poorly determined in individual fields
and the field-to-field dispersion in $r_0$ is poorly determined
since there are only four fields.  

Second, we broke our observed bright and faint catalogs into many
smaller sub-catalogs by rejecting a random fraction $p=0.5$--$0.9$
of the sources, observed how the dispersion in the best-fit $r_0$
value among the sub-catalogs depended on the number of objects in
the catalog, and extrapolated to the full catalog sizes.

We adopt compromise values for the standard deviation
in our final estimates for the $1\sigma$ confidence intervals
for the two analyses (eq.~\ref{eq:kpair} and~\ref{eq:kdaddi}) 
of the two catalogs,
$r_0 = 3.8\pm 0.8$, $4.4\pm 0.8 h^{-1}$ comoving Mpc for $K_s>20.5$ and
$r_0 = 10\pm 3$, $11\pm 8 h^{-1}$ comoving Mpc for $K_s\leq 20.5$.
The consistency of the answers for the two approaches suggests
that systematic problems are minimal.  We will adopt the 
the results from eq.~\ref{eq:kpair} for the remainder of
the paper, since that equation is subject to smaller random uncertainties.

To verify that the high value of $r_0$ for the bright sample
reflected genuine large-scale clustering, rather than (say)
the presence of merging sub-units within individual halos,
we repeated the analysis after excluding all pairs
with $\theta_{ij}<60''$ (i.e., projected separation $\simlt 1.1h^{-1}$ comoving
Mpc) from the calculation in equations~\ref{eq:kpair} and~\ref{eq:kdaddi}.
The best-fit values of $r_0$
were not significantly altered.

\section{DISCUSSION}
\label{sec:summary}

\begin{figure}
\plotone{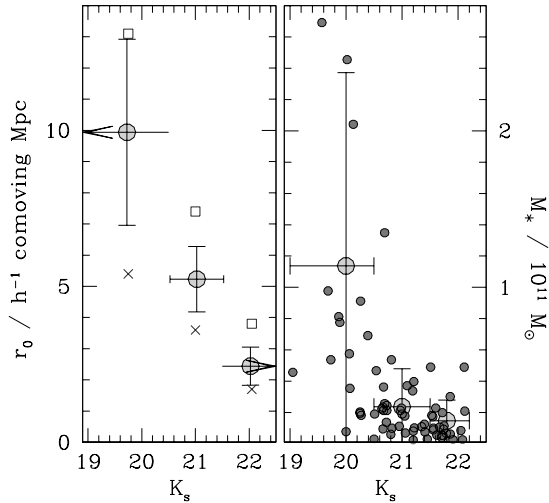}
\caption{
Left panel: correlation lengths for sub-samples with Vega-magnitude
$K_s\leq 20.5$, $20.5<K_s\leq 21.5$, and $K_s>21.5$.
The number of galaxies in the three subsamples is
75, 145, and 148, respectively.
The values shown are from equation~\ref{eq:kpair}.
Large circles are for the full sample;
their uncertainties in $r_0$ were estimated with the approach
of~\S~\ref{sec:results}. 
Small squares and crosses
show the results for Q1623 and for the other three fields,
respectively.
Right panel: Estimated stellar mass vs. $K_s$ magnitude for
galaxies in the sample of Steidel et al. (2005).
Small circles show the results for individual galaxies.
Large circles with error bars show the mean $\pm$ r.m.s. for
galaxies in different magnitude bins.
\label{fig:r0mstar_vs_K}
}
\end{figure}

Our analysis of the largest existing spectroscopic sample of galaxies
with $K_s<20.5$ at redshift $z\sim 2$ confirms previous reports
(Daddi et al. 2004) that these galaxies are strongly clustered.  
Their 
estimated correlation length of $r_0=10\pm 3h^{-1}$ Mpc
significantly exceeds that
of normal
galaxies at high or low redshift (e.g., Adelberger et al. 2005, Budavari et al. 2003).
The exact value of the correlation length at $K_s<20.5$ may be
more uncertain than this number implies,
since the Daddi et al. (2004) sample is
so small (see Adelberger 2005) and since much of our
signal comes from a field (Q1623; see figure~\ref{fig:r0mstar_vs_K}) which
contains an unusual concentration
of bright QSOs at $2.2\simlt z\simlt 2.6$\footnote{These QSOs are not
in the BX sample.  The sample analyzed here contains only 3 QSOs,
and their clustering does not significantly affect the results.
The point is that our surveyed volume may not be a fair sample
of the universe, since one of our fields contains one of the
richest known concentrations of bright QSOs.},
but the increase of clustering strength with $K_s$ luminosity
has been established with high significance ($>95$\%; see~\S~\ref{sec:results}).

The results of this paper, coupled with those of Shapley et al. (2004),
suggest that 
star-forming galaxies in optically-selected and near-IR-selected 
high-redshift surveys have similar properties
when the surveys are restricted to a common magnitude limit
$K_s\simlt 20.5$.  Galaxies with $K_s<20.5$
undoubtedly make up a larger fraction of the sources in
the near-IR surveys, 
but they do not seem to be
a fundamentally different population.
The left panel of figure~\ref{fig:r0mstar_vs_K} shows, for example, that their large
correlation lengths reflect the culmination of a trend
that is apparent
among the fainter galaxies
that dominate optical surveys.
Although existing optical surveys do not detect every star-forming
galaxy with $K_s<20.5$, the
fraction they find appears to be reasonably representative of the whole.

Since $K_s$ and ${\cal R}-K_s$ are correlated in our sample,
the relationship between $r_0$ and $K_s$ implies relationship between $r_0$
and ${\cal R}-K_s$.  We find that $r_0$ increases from $4.0\pm 0.8 h^{-1}$ Mpc
at ${\cal R}(AB)-K_s({\rm Vega})<3.5$ to $r_0=9\pm 3h^{-1}$ Mpc at ${\cal R}-K_s>3.5$.
A correlation of $r_0$ with both $K_s$ and ${\cal R}-K_s$ 
could result from an underlying correlation between $r_0$
and stellar mass $M_\ast$, since ${\cal R}-K_s$ and $K_s$ are both
correlated with $M_\ast$ (Shapley et al. 2005, in preparation).
An underlying correlation with $M_\ast$ could also explain
why $r_0$ varies so strongly with $K_s$.
Spitzer Space Telescope observations analyzed by Shapley et al. (2005, in preparation)
show that mass-to-light ratios in our sample increase significantly with
$K_s$ luminosity, which implies that $M_\ast$ is a strong function of $K_s$
(see the right panel of figure~\ref{fig:r0mstar_vs_K}).
The very high stellar masses associated with $K_s<20.5$ can only
be produced in unusually massive potential wells that formed
unusually early, and these are exactly the potential wells
that ought to be most strongly clustered.  The increase of
$r_0$ with $K_s$-luminosity may simply 
reflect the fact that galaxies in proto-cluster environments
formed earlier and were consequently able to convert more of
their baryons into stars by $z\sim 2$.  Steidel et al. (2005, in preparation)
discuss the point in more detail.

The strong dependence of galaxy properties on $K_s$ magnitude
can be viewed in two ways.  
On the one hand, since galaxies with $K_s\simlt 20$ are not representative
of the high-redshift population as a whole,
conclusions from bright near-IR-selected surveys
probably apply to only a limited subset of galaxies.
These galaxies are not irrelevant, since they are plausible progenitors
for the rarest and most massive ellipticals in the local universe,
but their properties should not be ascribed 
to the dominant fainter population.
On the other hand, the extreme properties of galaxies
with $K_s\simlt 20$ 
may help reveal trends that are difficult to discern
in fainter samples, and these trends 
could provide useful insight into the physics of galaxy formation.
Exploring the second possibility will require extensive
catalogs of high redshift galaxies with $K_s\simlt 20$.
The ability of optical surveys
to find them in large numbers is surely good news.

\bigskip
\bigskip
We thank the referee, A. Cimatti, for two detailed
reports.
KLA and AES were supported by
fellowships from the Carnegie Institute of Washington and the Miller Foundation.
DKE, NAR, and CCS were supported
by grant AST 03-07263 from the National Science Foundation
and by a grant from the Packard Foundation.
Section~\ref{sec:methods} was written in Port of Spain's 
Normandie Hotel, which KLA thanks for remarkable hospitality.

\end{document}